\documentclass[a4paper,showkeys, preprint,aps,10pt]{revtex4-1}

\usepackage{natbib}
\usepackage{graphics}
\usepackage{graphicx}
\usepackage{amsmath,amssymb}

\begin{document}

\title{Looking for the phase transition --- recent NA61/SHINE results}
\thanks{Lecture given at the Conference "Compact stars in the QCD Phase Diagram VI", in Dubna, September 26 - 29, 2017}
\author{Ludwik Turko}
\email{ ludwik.turko@ift.uni.wroc.pl}
\collaboration{for the NA61/SHINE Collaboration}
\affiliation{Institute of Theoretical Physics, University of Wroclaw, pl. M.~Borna 9, 50-205 Wroclaw, Poland}


\begin{abstract}
The fixed-target NA61/SHINE experiment at the CERN Super Proton Synchrotron (SPS) seeks to find the  critical point (CR) of strongly interacting matter as well as the properties of the onset of deconfinement. The experiment provides a scan of measurements of particle spectra and fluctuations in proton–proton, proton–nucleus, and nucleus–nucleus interactions as functions of collision energy and system size, corresponding to a two-dimensional phase diagram (T-$\mu_B$). New NA61/SHINE results are shown here, including transverse momentum and multiplicity fluctuations in Ar+Sc collisions as compared to NA61 p+p and Be+Be data, as well earlier NA49 A+A results.
Recently, a preliminary effect of change in the system size dependence, labelled as the "percolation threshold" or the "onset of fireball", was observed in NA61/SHINE data. This  effect  is  closely  related  to  the  vicinity  of  the  hadronic  phase  space transition region and will be discussed in the text.
\end{abstract}

\maketitle

%%%%%%%%%%%%%%%%%%%%%%%%%%%%%%%%%%%%%%%%%%
\section{Introduction}
\begin{figure}[!ht]
\centering
\includegraphics[width=\textwidth]{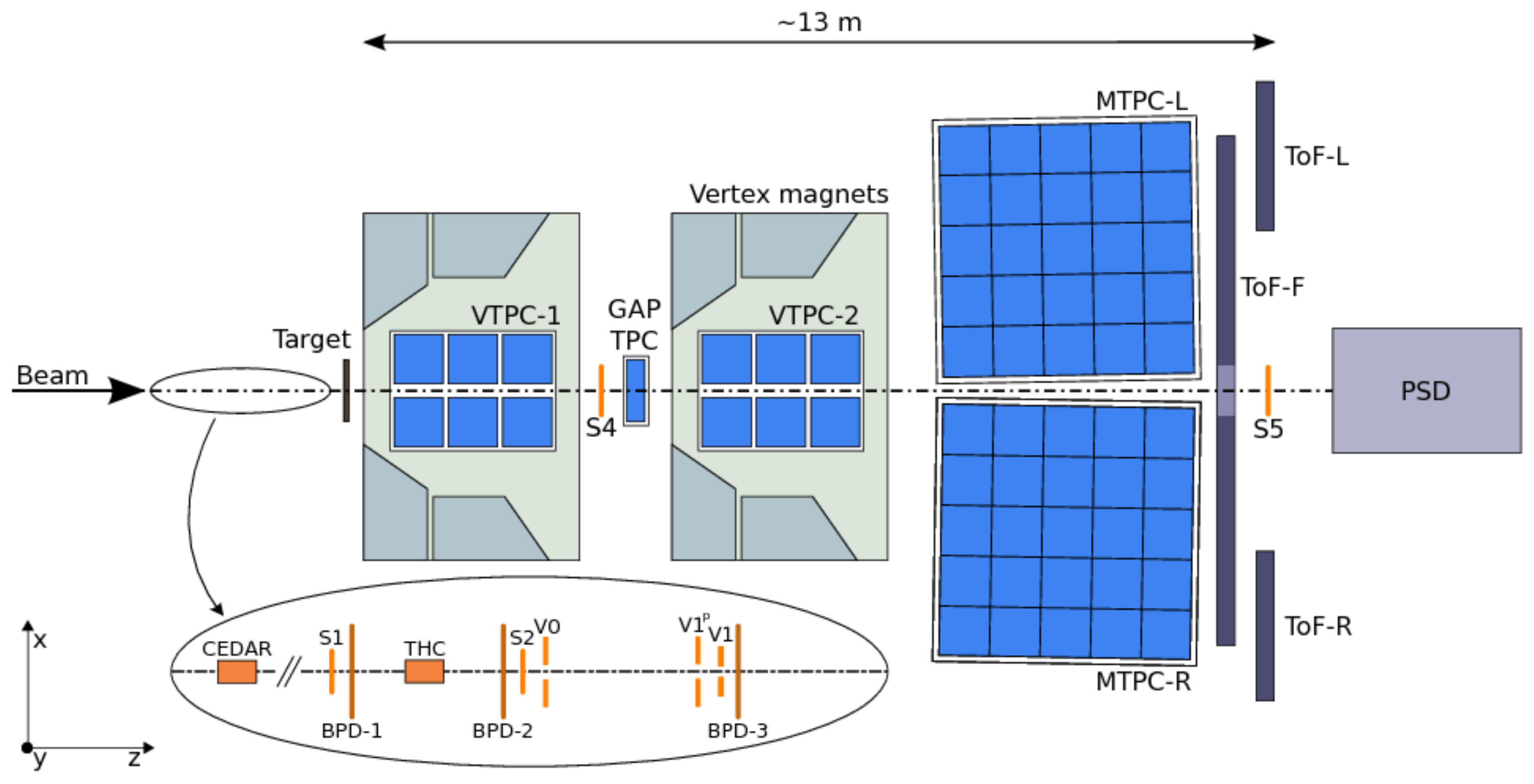}
\caption{The NA61/SHINE detector consists of a large acceptance hadron spectrometer followed by a set of six Time Projection Chambers (TPCs) as well as Time-of-Flight detectors (ToFs). The high resolution forward calorimeter, the Projectile Spectator Detector (PSD), measures energy flow around the beam direction. For hadron-nucleus interactions, the collision volume is determined by counting low momentum particles emitted from the nuclear target with the Low Momentum Particle Detector (a small TPC) surrounding the target. An array of beam detectors identifies beam particles, secondary hadrons and nuclei as well as primary nuclei, and measures their trajectories precisely.}\label{NA61}
\end{figure}

\begin{figure}[!ht]
%\centering
\includegraphics[width=0.5\textwidth]{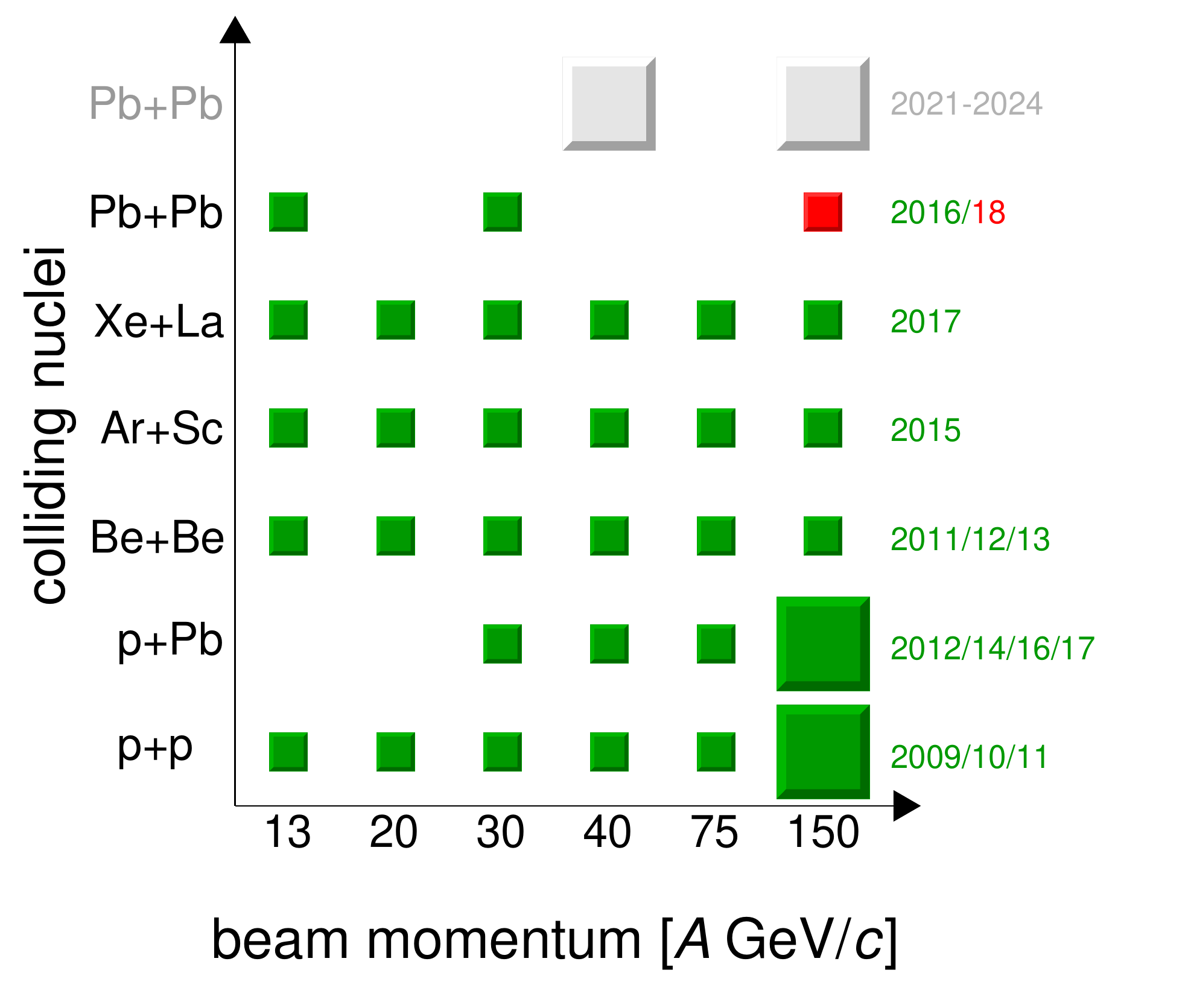}
\caption{For the programme on strong interactions, NA61/SHINE scans in the system size and beam momentum. In the plot, the recorded data are indicated in green, the approved future data taking in red, and the proposed extension for the period $\geqslant 2018$ in grey.}\label{beams}
\end{figure}

The NA61/SHINE, understood as the \textbf{S}uper Proton Synchrotron (SPS) \textbf{H}eavy \textbf{I}on and \textbf{N}eutrino \textbf{E}xperiment, is the continuation and the extension of the NA49 experiment \cite{Antoniou:2006_034,  Abgrall:2008_018}. It uses a similar experimental fixed-target setup to NA49 (Figure \ref{NA61})  but with an extended research programme. Beyond an enhanced strong interactions programme, measurements of hadron production for neutrino and cosmic ray experiments are realized. The collaboration involves about 150 physicists from 15 countries and 30 institutions. It is the second largest non-LHC (the Large Hadron Collider) experiment at the CERN.
\\
\\
The strong interaction programme of the NA61/SHINE is dedicated to the study of the onset of deconfinement and the search for  the critical point (CR) of hadronic matter, related to the phase transition between hadron gas (HG) and quark–gluon plasma (QGP). The NA49 experiment studied  hadron production in Pb+Pb interactions, while the NA61/SHINE collects data  varying beam energy within the range of 13A–-158A GeV and varying sizes of the colliding systems. This is equivalent to the two dimensional scan of the hadronic phase diagram in the $(T, \mu_B)$ plane, as depicted in Figure \ref{beams}. The ion collisions research programme was initiated in 2009 with the p+p collisions used later on as reference data  for heavy ion collisions.
\\
\\
Hadron-production measurements for neutrino experiments are just reference measurements of p+C interactions for the T2K experiment, since they are necessary for computing initial neutrino fluxes at J-PARC. It has been extended to measure the production of charged pions and kaons produced in interactions out of thin carbon targets and replicas of the T2K targets what is necessary to test accelerator neutrino beams \cite{Abgrall:2011ae}. Data taking began in 2007.

Collected p+C data also allow for better understanding of nuclear cascades in the {cosmic air} showers --- necessary in the Pierre Auger and KASCADE experiments \cite{Auger_2004,Kascade_2003}. These are reference measurements of p+C, p+p, $\pi$+C, and K+C interactions for  {cosmic ray} physics. {The cosmic ray collisions with the Earth's atmosphere  produce secondary air shower radiation. Some of particles produced in such collisions subsequently decay into muons, which are able to reach the surface of the Earth}.  Cosmic ray-induced muon production can allow the reproduction of primary cosmic ray composition if related hadronic interactions are known \cite{Morison_2008}.

\begin{figure}[!ht]
  \centering
  \includegraphics[width=0.5\textwidth]{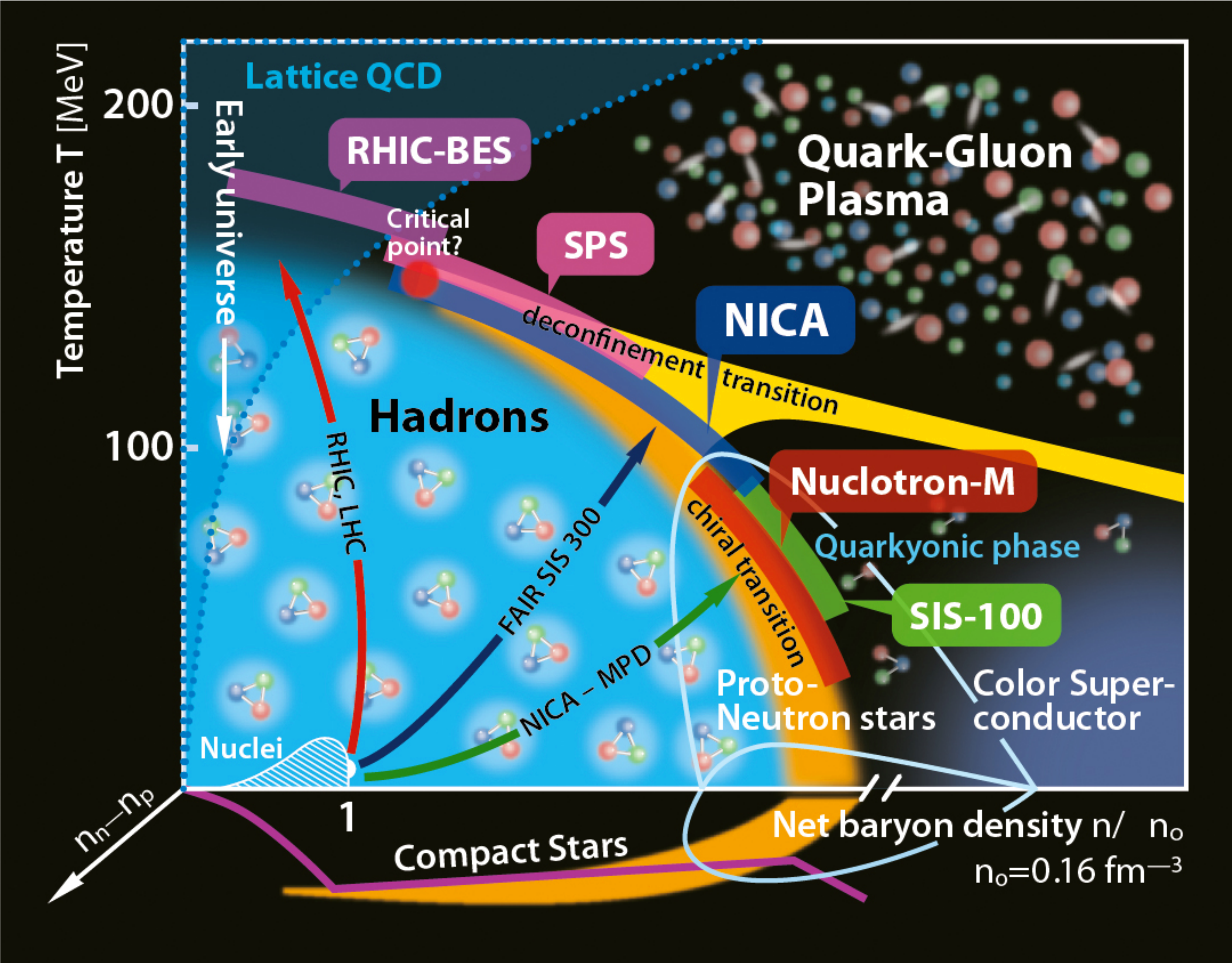}
  \caption{Phase diagram of strongly interacting matter in the temperature and baryonic chemical potential $( T,\mu_B)$ plane. Picture taken from this ($CSQCD\ 2017$) conference poster}
\label{phases}
\end{figure}

As seen in Figure \ref{phases}, the phase  structure  of  hadronic  matter  is involved. Progress in the theoretical understanding of the subject and the collection of more experimental data will allow us to delve further into the subject. While the highest energies achieved at the LHC and RHIC colliders provide data related to the crossover HG/QGP regions, the SPS fixed-target NA61/SHINE experiment is particularly suited to exploring the phase transition line of HG/QGP with the CR included.

Heavy ion collision (HIC) experiments reproduce the conditions of the first 10 $\mu$s
after the Big Bang, when a phase transition  from the QGP to a hadron gas would have taken place \cite{Satz_2013}. It appears, however, that the QGP can be present in the core of massive neutron stars --- particularly those with  masses exceeding two solar masses \cite{Benic:2014jia, Fischer:2017lag}. That would correspond to the far lower right part of the phase plot, beyond the $( T,\mu_B)$ range covered by Figure \ref{phases}. The CR has been long predicted  for thermal quantum chromodynamics (QCDs) at finite $\mu_B/T$ \cite{Barducci:1989wi,Halasz:1998qr, Berges:1998rc} although this was not unanimously accepted previously \cite{deForcrand:2008vr}. However, lattice QCD calculations are becoming more and more accurate, leading to the present conclusions that the cross-over region occurs at $T_c(\mu_B=0)=154\pm 9$ MeV  \cite{Bazavov:2014pvz} and the location of a CR is not expected for $\mu_B/T\leqslant 2$ and $T/T_c(\mu_B=0)>0.9 $ \cite{Bazavov:2017dus}. A more detailed exploration of QCD phase diagram would need both new experimental data with extended detection capabilities and improved theoretical models \cite{Caines:2017vvs}.

\begin{figure}[!ht]
  \centering
  \begin{tabular} {cc}
  \includegraphics[width=0.45\textwidth]{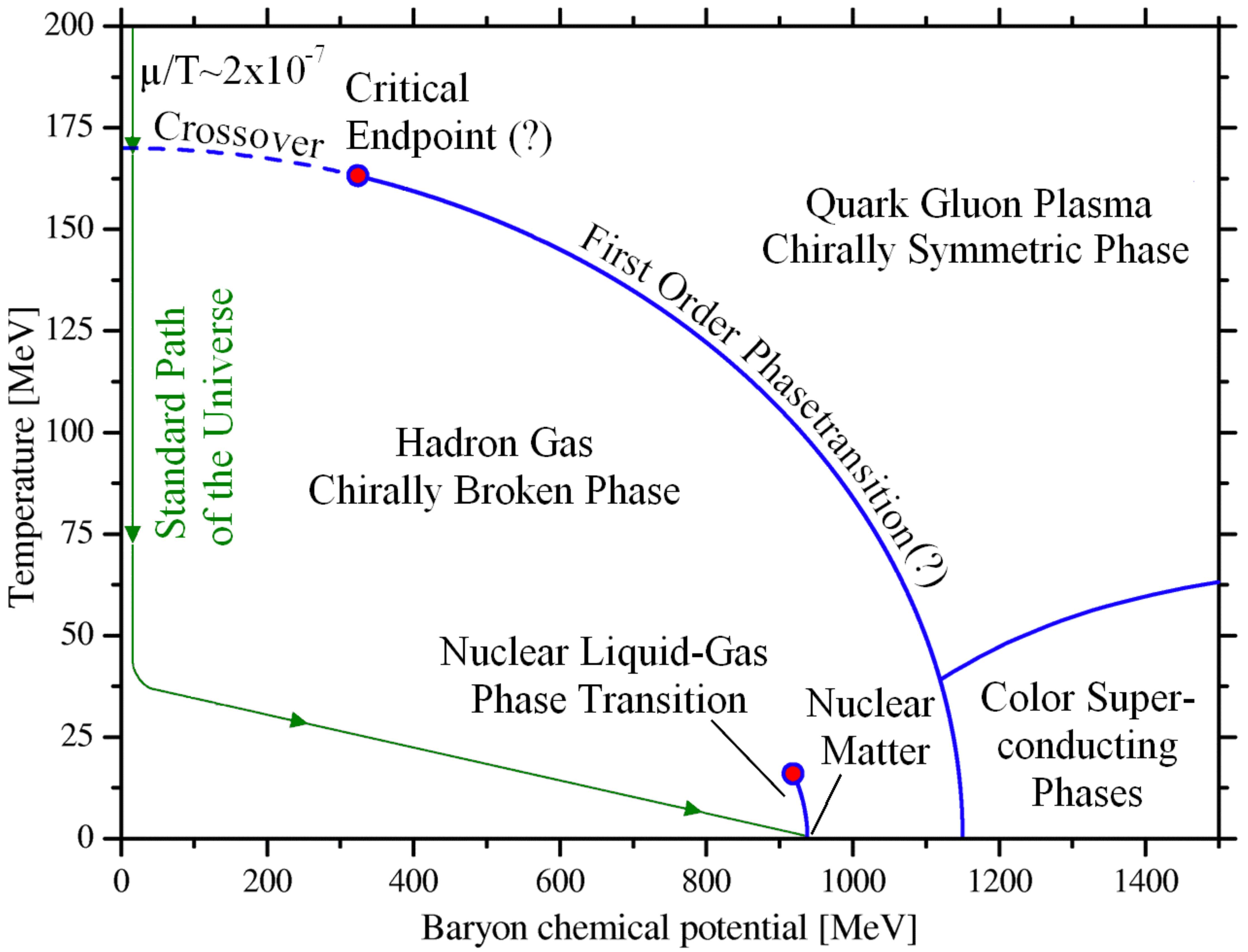}%\qquad
&\includegraphics[width=0.45\textwidth]{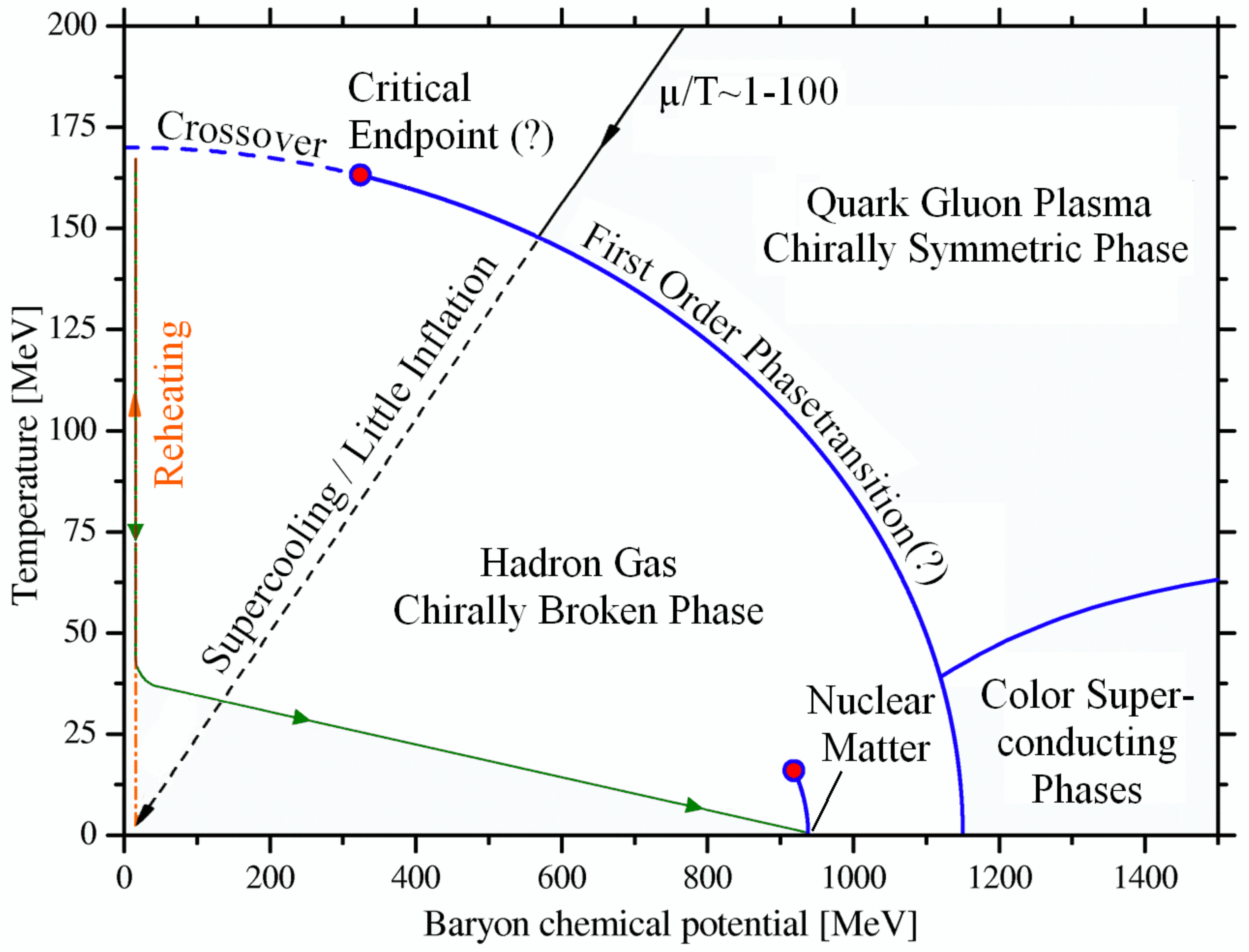}\\
({\bf a})&({\bf b})\\
\end{tabular}
  \caption{(\textbf{a}) Sketch of a possible quantum chromodynamic (QCD)-phase diagram with the commonly accepted standard evolution path of the universe as calculated e.g. in \cite{Fromerth:2002wb}. %\quad
 (\textbf{b}) Sketch of a possible QCD phase diagram with the evolution path in the scenario of the cosmic separation of~phases. }\label{LISc}
\end{figure}

Another intriguing and far reaching possibility is the Big Bang phase transition scenario, referred to by Edward Witten as the "cosmic separation of phases" \cite{Witten:1984rs}. In the standard approach, the Big Bang  QGP is almost matter-–antimatter symmetric and evolves to lower temperatures through the crossover region almost vertically to the temperature axis \cite{Fromerth:2002wb}. Edward Witten, using almost "back of envelope" arguments, pointed out the possibility of using the path of universe starting in the QGP phase from the high baryonic chemical potential region  {reaching} almost zeroth temperature, {with supercooled QGP}. Hadronization then becomes quite an explosive phenomena with  {a necessary subsequent reheating} (see e.g. \cite{Boeckel:2009ej,McInnes:2015hga}). Corresponding plots, taken from \cite{Boeckel:2009ej},  {are} shown in Figure \ref{LISc}.
\\

Beyond cosmological effects (little/tepid inflation) such a possibility would change also our understanding of the hadronization effect in HIC processes.

\section{New  NA61/SHINE Results}

\subsection{Irregularities --- the Horn}

It was expected \cite{Gazdzicki:1998vd} that ratio $K^+/\pi^+$ produced at HIC energies of about $\sqrt{s_{NN}}\thickapprox 10$ GeV should reach a rapid maximum when QGP formation begins. In 1998 there was not enough experimental data to fully check this hypothesis. Present collected results fully confirm the appearance of the horn, although there are still discussions about its relevance to the HG/QGP phase transition.

\begin{figure}[!ht]
\centering
\includegraphics[width=0.5\textwidth]{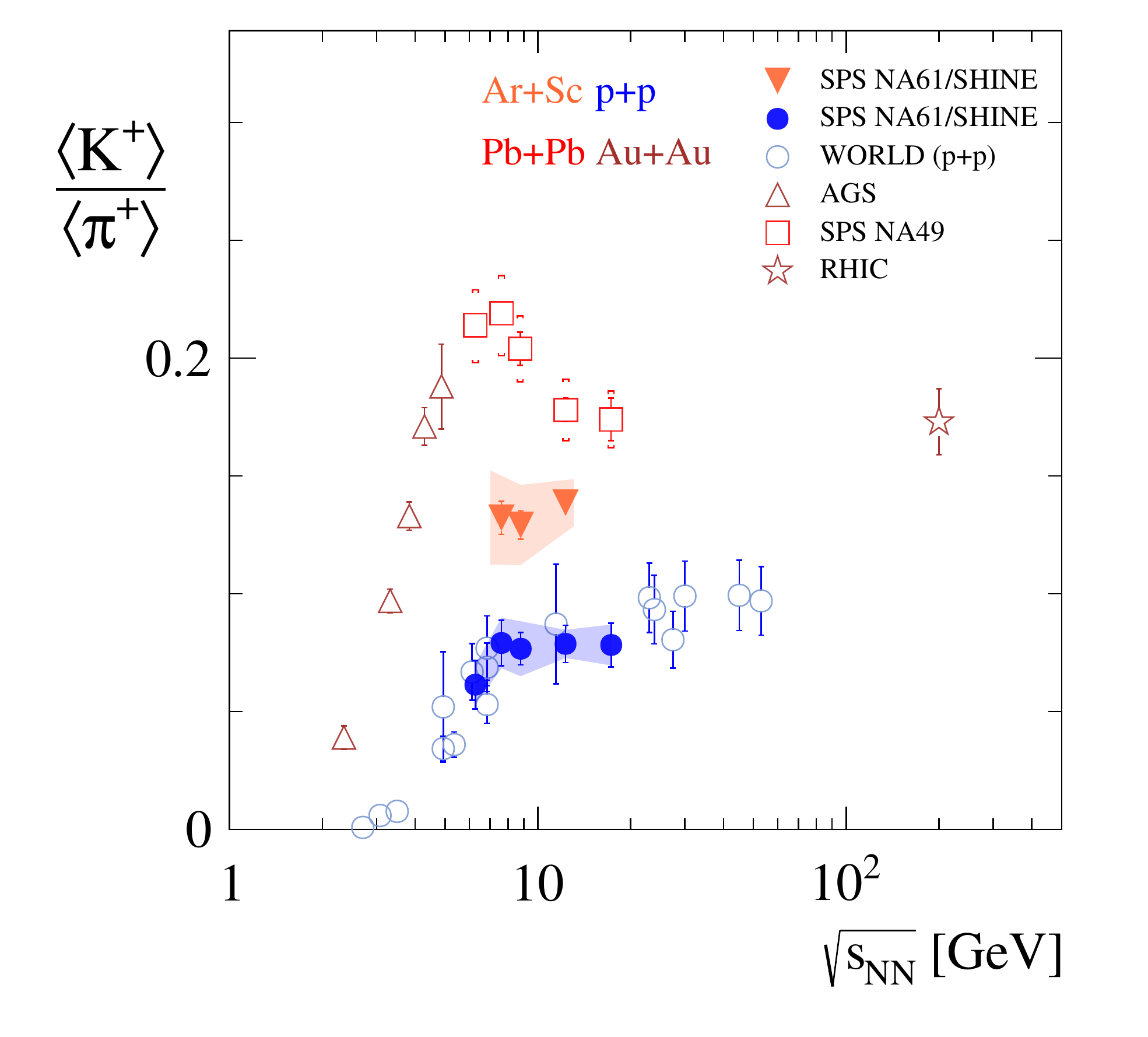}
\caption{{Horn:} a strong maximum of the ratio of $K^+/\pi^+$ multiplicities. A reduced shadow of the horn structure is visible in  {p+p} reactions.}\label{horn}
\end{figure}

Recent data from NA61/SHINE \cite{Lewicki:2017ueb} show also {a} strong dependence of the effect on the size of the colliding  objects, as seen in Figure \ref{horn}.
\\

\subsection{ Irregularities --- the Step}

\begin{figure}[!ht]
\centering
\centering
\begin{tabular}{ccc}
\includegraphics[width=0.35\textwidth]{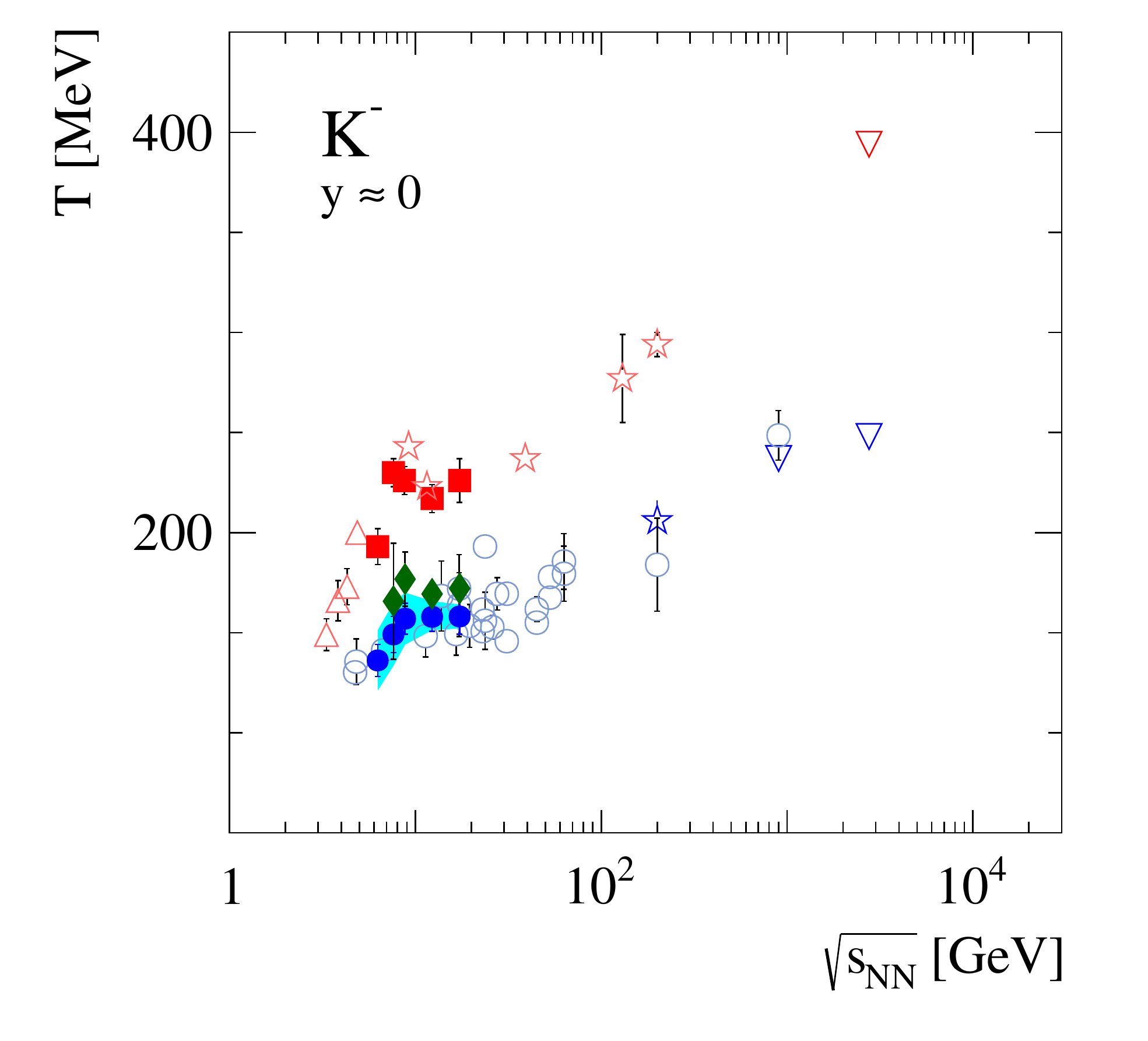} %\qquad
&\includegraphics[width=0.35\textwidth]{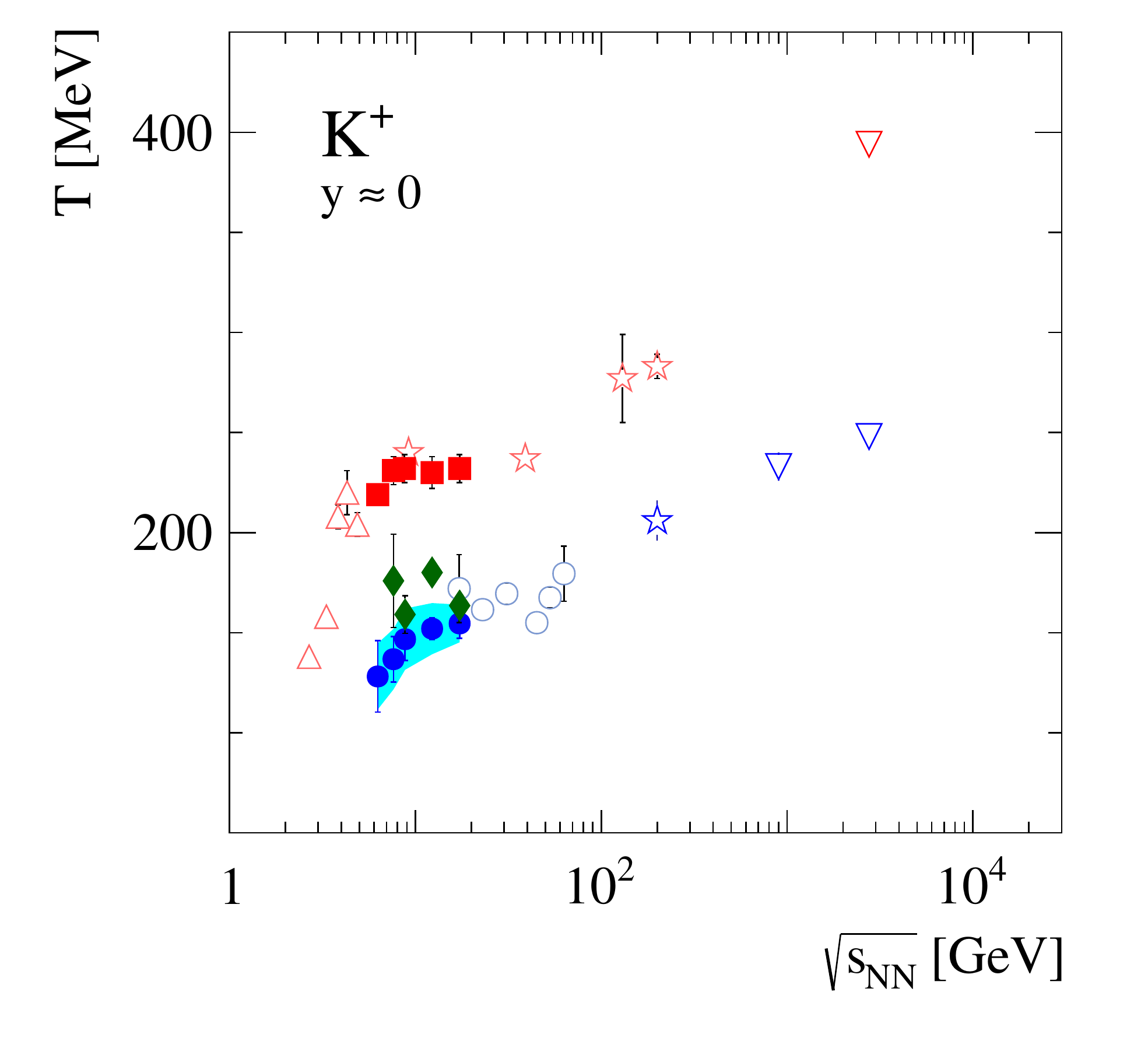}
&\includegraphics[width=0.20\textwidth]{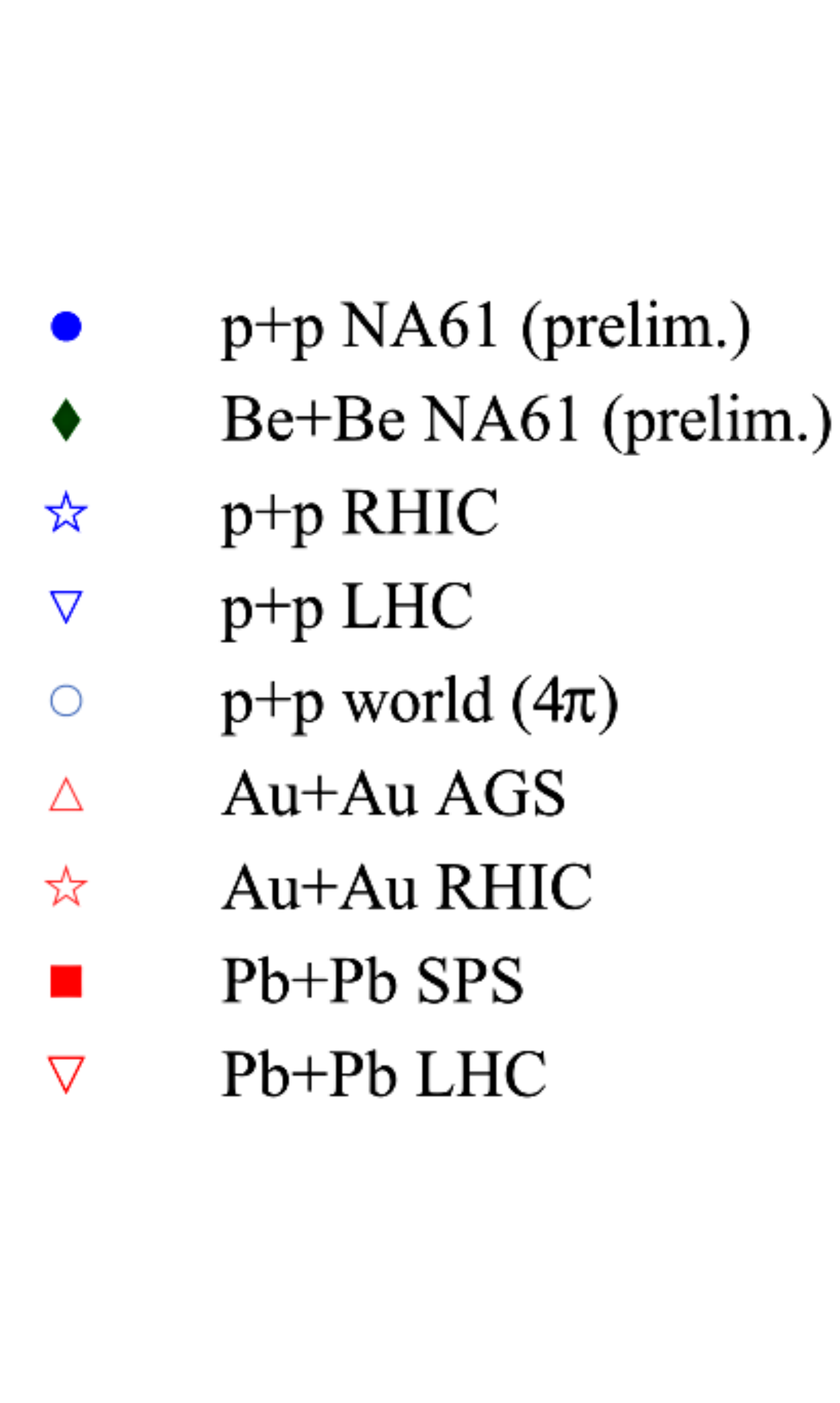}\\
({\bf a})&({\bf b})&({\bf c})\\
\end{tabular}
\caption{Inverse slope parameters $T$ of negative ({\bf a}) and positive ({\bf b}) kaons exhibit rapid changes in the SPS energy range --- also seen in  {p+p}~collision. Data collected from all available energy ranges ({\bf c}).}  \label{Fig.6}
\end{figure}

Plateau: A step-in the inverse slope parameter $T$ of the transverse mass spectra $m_T$ at mid-rapidity ($0 < y < 0.2$) plotted against the collision energy per nucleon (Figure \ref{Fig.6}) is expected for the onset of deconfinement \cite{Gazdzicki:1998vd}. The effect increases with the size of colliding objects. Qualitatively, a similar structure is visible in p+p collisions, with Be+Be slightly above, consistent with the step structure.

\subsection{Fluctuations}

Experimental facilities of NA61  allows to measure fluctuations of various physical quantities which are sensitive to the vicinity of the critical point. These fluctuations can create a signature for the critical point. An analysis of  fluctuations of various observables, particularly in a range of energies around 8 GeV per colliding nucleon pair at the center of mass in interactions of light nuclei  (it corresponds to the beam energy of 30 GeV in the frame of a stationary target), is the main goal of the NA61/SHINE experiment. This is just the kinematical region where NA49 data indicate that the onset of deconfinement in central Pb+Pb collisions. It is mainly based on the observation of structures in the energy dependence of hadron production in central Pb+Pb collisions which are not observed in elementary interactions \cite{Alt:2008,Gazdzicki:2008}

\begin{figure}[!ht]
\begin{tabular} {cc}
\includegraphics[width=0.45\textwidth]{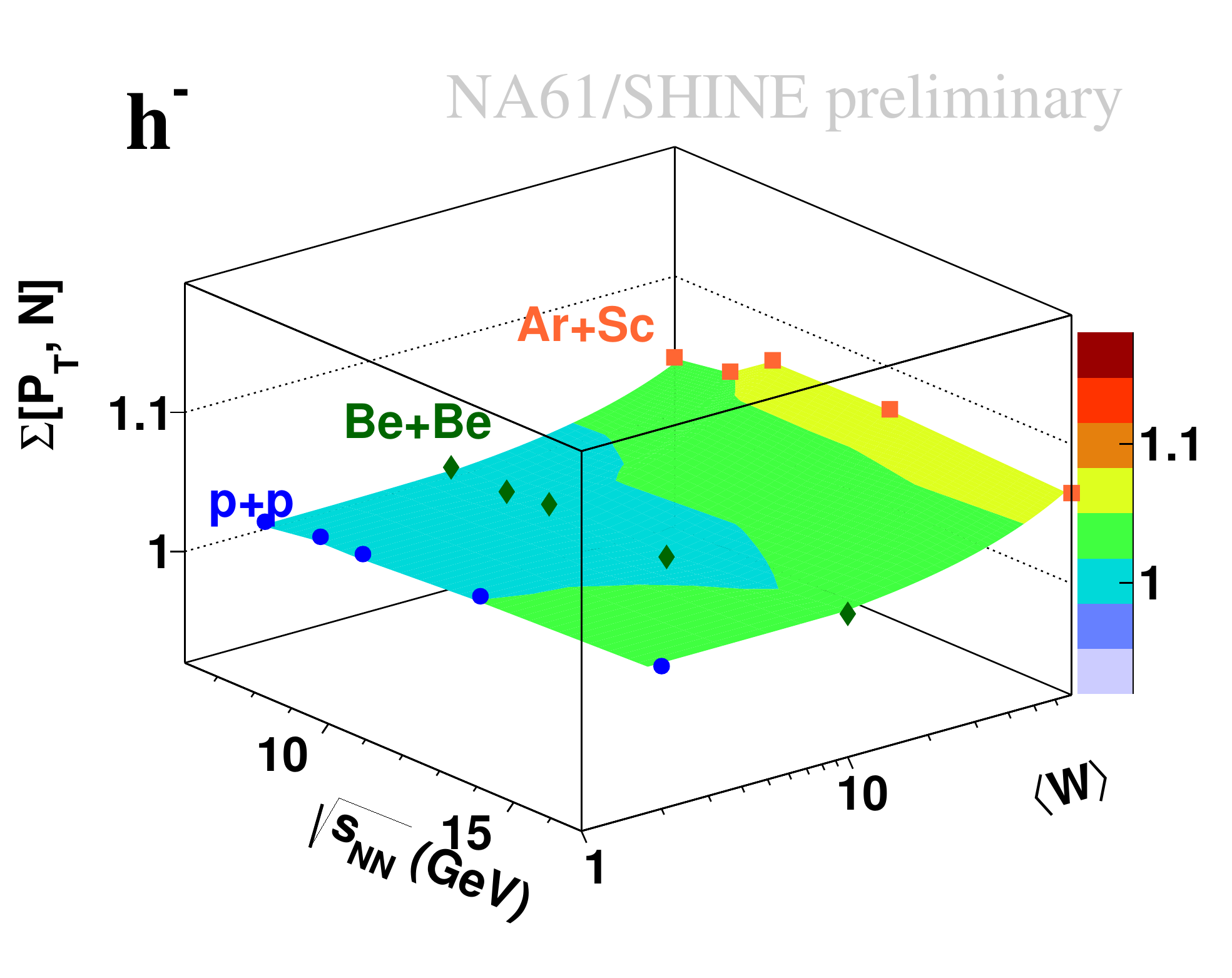}\qquad
&\includegraphics[width=0.45\textwidth]{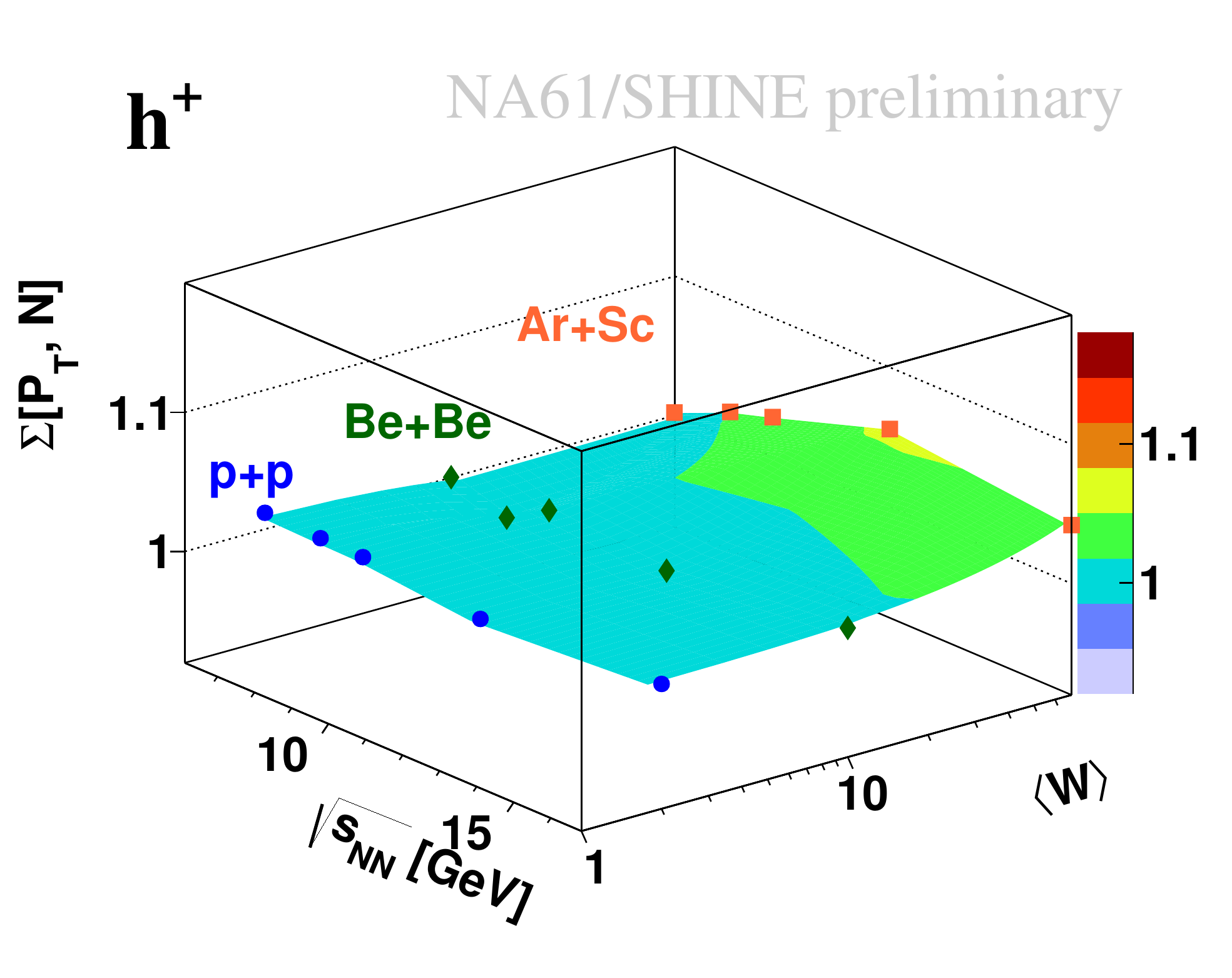}\\
({\bf a})&({\bf b})
\end{tabular}
\caption{
 Critical fluctuations in $p_T$ of negative (\textbf{a})  and positive (\textbf{b}) charged hadrons in  ${}^{40}\mbox{Ar +} {}^{45}\mbox{Sc}$, ${}^7\mbox{Be}+{}^9\mbox{Be}$ and {p+p} collisions. }\label{fluct}
\end{figure}

Preliminary data presented in Figure \ref{fluct} do  not show any signs of critical behaviour \cite{Aduszkiewicz:2017mei}.

\section{System size dependence}

In the recent months, some unexpected results were observed by the NA61/SHINE~experiment \cite{Lewicki:2017ueb,Seryakov:2017jpr,Gazdzicki:2017zrq}{{, concerning} qualitative differences in system size dependence behaviour. It appears that in particular Be+Be results are very close to p+p at different collision energies. {An example} of such behaviour is presented in Figure \ref{MFBe}.

\begin{figure}[!ht]
  \centering
\begin{tabular} {cc}
  \includegraphics[width=0.4\textwidth]{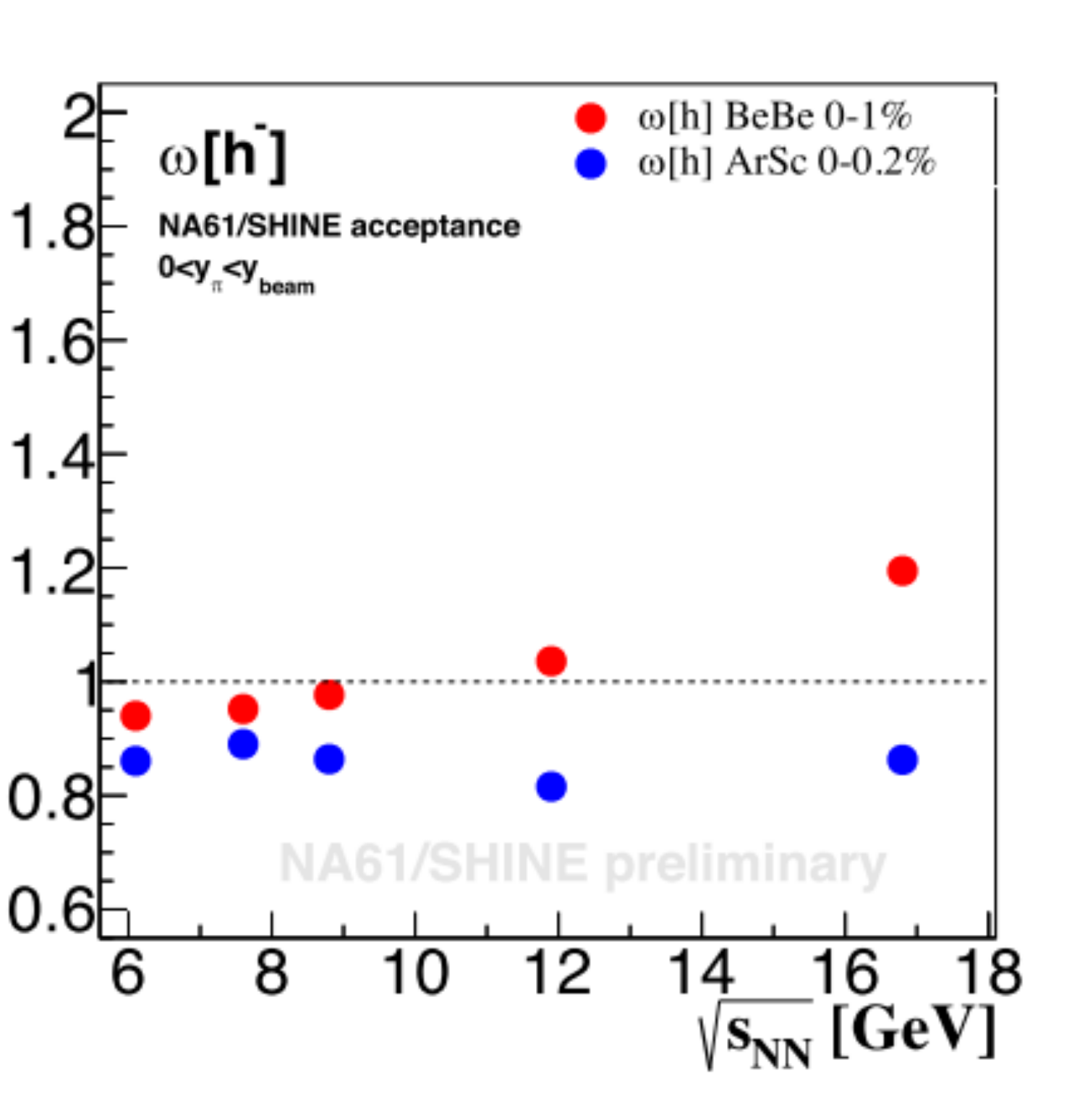}
 & \includegraphics[width=0.4\textwidth]{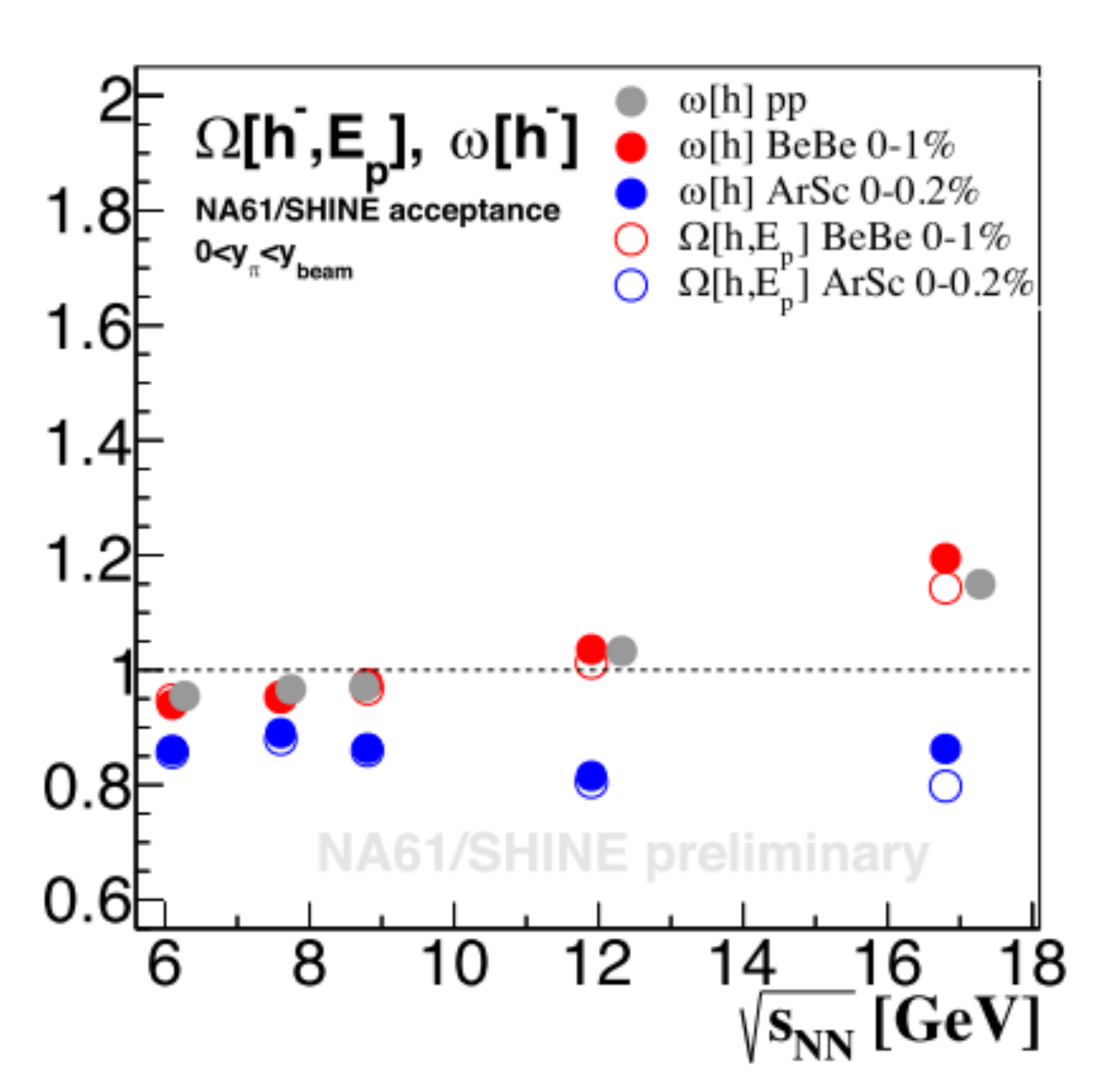}\\
({\bf a})&({\bf b})\\
\end{tabular}

  \caption{(\textbf{(a)})  multiplicity fluctuation increases with collision	
energy	in	{Be+Be} but remains constant in {Ar+Sc}.\
(\textbf{b})  multiplicity fluctuation in  {$\mbox{Ar + Sc}, {}^7\mbox{Be} + {}^9\mbox{Be}$ and p+p} collisions.  {Be+Be} almost identical to	 {p+p} fluctuation {within statistical errors given by plot's points sizes.} }\label{MFBe}
\end{figure}

It looks as if with the increasing size of colliding systems, light clusters are produced more and more copiously, and at some density they start to overlap to reach percolation threshold.  This effect would not depend on energy  only, but also on the size of the system.

Intensive work now is ongoing to achieve more conclusive results, analyzing recently collected data from Xe+La and Pb+Pb.

\section{Conclusions}

The Holy Grail of HIC --- Quark Gluon Plasma --- still remains an elusive object.  Although there are no at present discussions  concerning the very existence of this state of matter, there are still open problems connected with experimental signatures of many theoretical ideas and predictions in this field. The NA61/SHINE experiment acts in the energy region particularly suited for the appearance of phase transition effects. Beyond this, the fixed-target technology makes possible to perform $4\pi$ geometry measurementswhich are not accessible  in collider-type experiments.

To date, collected and analyzed NA61/SHINE data related to theoretical predictions of fluctuations in the presence of CR have not shown any anomaly that could be attributed to this. These data relate to  $N-p_T$ fluctuations in p+p, Be+Be, and Ar+Sc central events.

There is clear system size dependence of $m_T$ spectra that differs significantly between p+p and A+A events. This is the effect associated to the transverse collective flow.

The appearance of horn (Figure \ref{horn}) and step (Figure \ref{Fig.6}) is in accordance with theoretical predictions for the onset of deconfinement in HIC due to mixed phase of HG and QGP \cite{Gazdzicki:1998vd}.

The recent preliminary results of the NA61/SHINE concerning system size dependence may be also a signature for the new physical phenomena. There is a clearly visible jump between light and heavy systems. Be+Be results are very close to the p+p results, independently on the collision energy. In addition, multiplicity fluctuations, close to p+p value in Be+Be collisions, are strongly suppressed in Ar+Sc collisions.

For the CERN long shutdown 2019-2020, an NA61/SHINE detector upgrade system is foreseen. This would make the precise measurements of open charm and multi-strange hyperon production possible, which are also of great importance both for the neutrino physics programme as well as for the  precision measurements of cosmic rays.

\begin{acknowledgments}
The author acknowledges support by the Polish National Science Center under contract No. UMO-2014/15/B/ST2/03752 and by the Bogoliubov-Infeld programme for scientific collaboration between JINR Dubna and Polish Institutions.
\end{acknowledgments}

\appendix
\section{Abbreviations} The following abbreviations are used in this manuscript:\\

\noindent
\begin{tabular}{@{}ll}
AGS & Argonne National Laboratory\\
CERN & Conseil Europ{\'e}n pour la Recherche Nucl{\'e}aire\\
CR & critical point\\
HG & hadron gas\\
J-PARC & Japan Proton Accelerator Research Complex\\
LHC & Large Hadron Collider\\
HIC & heavy ion collision\\
QCD & quantum chromodynamics\\
QGP & quark-gluon plasma\\
RHIC & Relativistic Heavy Ion Collider\\
SPS & Super Proton Synchrotron
\end{tabular}

%%%%%%%%%%%%%%%%%%%%%%%%%%%%%%%%%%%%%%%%%%

\end{document}